\documentclass[epj]{svjour}
\usepackage{amssymb}
\usepackage{graphicx}
\usepackage{bm}
\usepackage{epstopdf}
\usepackage{epsfig}
\usepackage{color,framed}
\usepackage{amsmath,amssymb}
\begin{document}


\title{Localization of five-dimensional Elko Spinors on dS/AdS Thick Branes}


\author{Xiang-Nan Zhou$^{1}$,
        Xin-Yuan Ma$^{1}$,
        Zhen-Hua Zhao$^{2}$,
       and Yun-Zhi Du$^{3}$}                     
%
\mail{duyzh13@lzu.edu.cn}
\institute{\small$^1$College of Physics and Information Engineering, Shanxi Normal University, Linfen 041004, People's Republic of China\\
  \small$^2$Department of Applied Physics, Shandong University of Science and Technology, Qingdao 266590, People's Republic of China\\
  \small$^3$Institute of Theoretical Physics, Datong University, Datong 037009, People's Republic of China}
\date{Received: date / Revised version: date}
%
\abstract{Different from the Dirac spinor, the localization of a five-dimensional Elko spinor on a brane with codimension one is very difficult {because of} the special structure of the Elko spinor. By introducing the coupling between the Elko spinor and the scalar field generating the brane, we have two mechanisms for localization of the zero mode of a five-dimensional Elko spinor on a brane. One is the Yukawa-type coupling and the other is the non-minimal coupling. In this paper, we investigate the localization of the Elko zero mode on de Sitter and Anti-de Sitter thick branes with the two localization mechanisms, respectively. It shows that both the mechanisms can realize the localization. The forms of the coupling functions of the scalar field in the two mechanisms have similar properties respectively and they play similar role for the localization.
\PACS{
      {11.27.+d}{Extended classical solutions; cosmic strings, domain walls, texture}\and
     {04.50.-h}{Higher-dimensional gravity and other theories of gravity}
     }
} 
\maketitle

\section{Introduction}
\label{section1}
There are many classical problems which the Standard Model (SM) can not interpret sufficiently, such as the hierarchy problem \cite{ArkaniHamed:1998rs,Randall:1999ee,Antoniadis:1998ig,Das:2007qn,Yang:2012dd,Guo:2015wua}, cosmological problem \cite{ArkaniHamed:2000eg,Kim:2000mc,Starkman:2000dy,Starkman:2001xu,Kehagias:2004fb,George:2008vu,Dey:2009xu,Haghani:2012zq} and the nature of dark matter and dark energy \cite{ArkaniHamed:1998nn,Sahni:2002dx,Cembranos:2003mr,Nihei:2004xv}. Because extra dimension and brane-world theories can provide new mechanisms to solve these problems, they have attracted more and more attention since the famous Arkani-Hamed, Dimopoulos and Dvali (ADD) \cite{ArkaniHamed:1998rs} and Randall-Sundrum (RS) brane-world models \cite{Randall:1999ee,Randall:1999vf} were presented. Unlike early models where the brane is a geometric hypersurface embedded in a higher dimensional spacetime, in a more realistic brane model the brane should have thickness and inner structure. Such thick branes can be generated by bulk matter fields (mostly scalar fields) \cite{Gremm:1999pj,DeWolfe:1999cp,Kobayashi:2001jd,Bazeia:2002sd,Wang:2002pka,Bazeia:2004dh,Afonso:2006gi,Bazeia:2006ef,Bogdanos:2006qw,Dzhunushaliev:2009va,Liu:2009ega,Liu:2009dt,Liu:2011wi,Guo:2011wr,Liu:2012rc,Liu:2012gv,Bazeia:2012br,German:2013sk,Dutra:2014xla,Guo:2014bxa,Geng:2015kvs}, or can be realized in pure gravities \cite{BarbosaCendejas:2005kn,HerreraAguilar:2010kt,Zhong:2015pta} (see Refs. \cite{Dzhunushaliev:2009va,Liu:2017} for more
detailed introduction about thick brane models).

In the braneworld scenario, an important and interesting issue is to investigate the mechanism by which the  Kaluza-Klein (KK) modes of various fields could be localized on the brane. These KK modes contain the information of extra dimensions. Especially, the zero modes of matter fields on the brane stand for the four-dimensional massless particles, and they can rebuild the four-dimensional SM at low energy. A lot of work about the localization of various matter fields on branes have been done \cite{Guo:2011qt,Bajc:1999mh,Oda:2000zc,Liu:2007ku,Liu:2008wd,Liu:2009uca,Guerrero:2009ac,Zhao:2010mk,Chumbes:2010xg,Chumbes:2011zt,Germani:2011cv,Liu:2011zy,Cruz:2012kd,Fu:2013ita,Liu:2013kxz,Zhao:2014gka,Alencar:2014moa,Vaquera-Araujo:2014tia,Du:2015pjw,Fu:2015cfa,Fu:2016vaj,Zhang:2016ksq,Li:2017dkw,Zhou:2017xaq}.

On the other hand, Elko spinor, which is named from the eigenspinor of the charge conjugation operator, has attracted continuing interest since it was introduced by Ahluwalia and Grumiller in 2005 \cite{Ahluwalia:2004sz,Ahluwalia:2004ab}. It is a new spin-1/2 quantum field which satisfies the Klein-Gordon (KG) equation rather than the Dirac one, and only interacts with itself, Higg fields and gravity \cite{Ahluwalia:2004sz,Ahluwalia:2004ab,Ahluwalia:2008xi,Ahluwalia:2009rh,Dias:2010aa,Lee:2012td,Lee:2015jpa,Fabbri:2010va}. As a candidate of dark matter, it was widely investigated in particle physics \cite{Ahluwalia:2004sz,Ahluwalia:2004ab,Ahluwalia:2008xi,Ahluwalia:2009rh}, cosmology \cite{Boehmer:2008rz,Boehmer:2008ah,Boehmer:2010ma,Wei:2010ad,Gredat:2008qf,Basak:2014qea,Pereira:2014wta,Pereira:2016emd,Pereira:2017efk,Pereira:2017bvq} and mathematical physics \cite{HoffdaSilva:2009is,HoffdaSilva:2016ffx,daRocha:2011yr,Ablamowicz:2014rpa,Rogerio:2016grn,Cavalcanti:2014wia,Cavalcanti:2014uta}.

In Refs. \cite{Liu:2011nb,Jardim:2014cya,Jardim:2014xla}, the localization of the zero mode of a five-dimensional Elko spinor on various Minkowski branes has been considered. A coupling mechanism should be introduced in order to localize the Elko zero mode on a brane. The first choice is the Yukawa-type coupling $-\eta F(\phi$ or $R)\mathop \lambda\limits^\neg\lambda$ between the five-dimensional Elko spinor $\lambda$ and the background scalar field $\phi$ \cite{Liu:2011nb,Jardim:2014cya} or the Ricci scalar $R$ \cite{Jardim:2014xla}. Here $F$ is a function of the background scalar field or the Ricci scalar and $\eta$ is the coupling constant. Recently, another localization mechanism, the non-minimal coupling $f(\phi)\mathfrak{L}_{Elko}$ between the Elko spinor and the background scalar field, has been investigated in Ref. \cite{Zhou:2017bbj}. Here $\mathfrak{L}_{Elko}$ is the Lagrangian of the Elko spinor and $f(\phi)$ is a function of the background scalar field. It has been shown that by introducing an auxiliary function $K(z)$, the general expression of the Elko zero mode $\alpha_0$ and the scalar function $f(\phi)$ could be obtained. And different forms of $K(z)$ will lead to different solutions of the zero mode and the scalar function. Thus, a non-minimal coupling can also provide the possibility of localizing the Elko zero mode. By now, all of the investigations about the localization of a five-dimensional Elko spinor were concerned with Minkowski branes. As we know, the properties of de Sitter (dS) and Anti-de Sitter (AdS) branes are very different from those of Minkowski branes, and the results of the localization of matter fields on dS and AdS branes are different compared to those on Minkowski branes. Thus, the localization of a five-dimensional Elko spinor on dS and AdS branes is an interesting topic. {At the same time, there are two kinds of localization mechanisms. What are the differences and similarities between them also attract our great interest. We believe that investigating the differences and similarities between them will be helpful to further explore the new localization mechanism and expand the possibility of localizing fields. Therefore,} in this paper, we will investigate this localization problem with the above two mentioned mechanisms. It will be shown that both localization mechanisms will work for dS and AdS thick brane models and the functions {$F(\phi)$ and $f(\phi)$} play a similar role for the localization.

This paper is organized as follows. We first review {the Yukawa-type and non-minimal couplings in Sec.~\ref{section2}. The localization of the zero mode of a five-dimensional Elko spinor on the dS and AdS thick branes is investigated with the two mechanisms in Sec.~\ref{section3}. Then in Sec.~\ref{section4}, we consider the localization of the Elko zero mode on another AdS thick brane.} Finally, a brief conclusion is given in Sec.~\ref{section5}.

\section{Review of localization mechanisms}
\label{section2}

In this section, we review the two localization mechanisms {for a five-dimensional Elko spinor in a thick brane model, namely, the Yukawa-like coupling and the non-minimal coupling between the Elko spinor and the background scalar field generating the thick brane}.

The line-element {is} generally assumed as   
\begin{eqnarray}
ds^{2}=\text{e}^{2A(y)}\hat{g}_{\mu\nu}dx^{\mu}dx^{\nu}+dy^{2}, \label{the line-element for 5D space time}
\end{eqnarray}
where {the warp factor $\text{e}^{2A(y)}$ is a function of the extra dimension $y$. By} performing the following coordinate transformation
\begin{eqnarray}
dz=\text{e}^{-A(y)}dy, \label{coordinate transformation}
\end{eqnarray}
the metric (\ref{the line-element for 5D space time}) {is transformed as}
\begin{eqnarray}
ds^{2}=\text{e}^{2A}(\hat{g}_{\mu\nu}dx^{\mu}dx^{\nu}+dz^{2}), \label{conformally flat line-element}
\end{eqnarray}
which is more convenient for discussing the {localization} of gravity and various matter fields.

\subsection{Yukawa-like coupling}
Firstly, we {start with the action of a five-dimensional massless Elko spinor}
\begin{eqnarray}
S&=&\int d^5x\sqrt{-g}\mathfrak{L}_{Elko},\nonumber \\
\mathfrak{L}_{\text{Elko}}&=&-\frac{1}{4}g^{MN}\left(\mathfrak{D}_{M}\mathop \lambda\limits^\neg\mathfrak{D}_{N}\lambda+\mathfrak{D}_{N}\mathop \lambda\limits^\neg\mathfrak{D}_{M}\lambda\right)-\eta F(\phi)\mathop \lambda\limits^\neg\lambda,~~~~\label{Lagrangian for Elko in 5D Yukawa}
\end{eqnarray}
{where the last term is the Yukawa-like coupling with $F(\phi)$ a function of the background scalar field $\phi$, and $\eta$ is the coupling constant.} In this paper, $M,N,\cdots=0,1,2,3,5$ and $\mu,\nu,\cdots=0,1,2,3$ denote the five-dimensional and four-dimensional spacetime indices, respectively. The covariant derivatives are
\begin{eqnarray}
\mathfrak{D}_{M}\lambda=(\partial_{M}+\Omega_{M})\lambda,~~\mathfrak{D}_{M}\mathop \lambda\limits^\neg=\partial_{M}\mathop \lambda\limits^\neg-\mathop \lambda\limits^\neg\Omega_{M}, \label{covariant derivatives}
\end{eqnarray}
where the
{spin} connection $\Omega_{M}$ is defined as
\begin{eqnarray}
   \Omega_{M}&=&-\frac{i}{2}\left(e_{\bar{A}P}e_{\bar{B}}^{~~N} \Gamma^{P}_{MN}
                             -e_{\bar{B}}^{~~N}\partial_{M}e_{\bar{A}N}\right)
                  S^{\bar{A}\bar{B}},\nonumber \\
   S^{\bar{A}\bar{B}}&=&\frac{i}{4}[\gamma^{\bar{A}},\gamma^{\bar{B}}].
    \label{tangent space connection}
\end{eqnarray}
Here $e^{\bar{A}}_{~M}$ is the vierbein and satisfies the orthonormality relation $g_{MN}=e^{\bar{A}}_{~M}e^{\bar{B}}_{~N}\eta_{\bar{A}\bar{B}}$. {$\bar{A},\bar{B}\cdots=0,1,2,3,5$ stand for} the five-dimensional local Lorentz indices. So the non-vanishing components of the spin {connection $\Omega_{M}$ are}
\begin{eqnarray}
\Omega_{\mu}=\frac{1}{2}\partial_{z}A\gamma_{\mu}\gamma_{5} +  \hat{\Omega}_{\mu} . \label{spin connection for brane}
\end{eqnarray}
Here {$\gamma_{\mu}$ and $\gamma_{5}$ are} the {four-dimensional} gamma matrixes on the brane, {and they} satisfy $\{\gamma^{\mu},\gamma^{\nu}\}=2\hat{g}^{\mu\nu}$.

Then the equation of motion for the Elko {spinor} coupled with the scalar {field} is read as
\begin{eqnarray}
\frac{1}{\sqrt{-g}}\mathfrak{D}_{M}(\sqrt{-g}g^{MN}\mathfrak{D}_{N}\lambda)-2\eta F(\phi)\lambda=0. \label{Elko's motion equation 2 Yukawa}
\end{eqnarray}
By considering {the metric} (\ref{conformally flat line-element}) and using the non-vanishing components of the spin connection (\ref{spin connection for brane}), the above equation can be rewritten as
 \begin{eqnarray}
  && \frac{1}{\sqrt{-\hat{g}}}\hat{\mathfrak{D}}_{\mu}(\sqrt{-\hat{g}}\hat{g}^{\mu\nu}\hat{\mathfrak{D}}_{\nu}\lambda)
 +\bigg[-\frac{1}{4}{A'}^{2}\hat{g}^{\mu\nu}\gamma_{\mu}\gamma_{\nu}\lambda\nonumber\\
 &+&\frac{1}{2}A'
 \Big( \frac{1}{\sqrt{-\hat{g}}} \hat{\mathfrak{D}}_{\mu}
       ( {\sqrt{-\hat{g}}} \hat{g}^{\mu\nu}\gamma_{\nu}\gamma_{5}\lambda)
 +\hat{g}^{\mu\nu}\gamma_{\mu}\gamma_{5}\hat{\mathfrak{D}}_{\nu}\lambda\Big)
 \nonumber\\
 &+&\text{e}^{-3A}\partial_{z}(\text{e}^{3A}\partial_{z}\lambda)\bigg]-2\eta \text{e}^{2A}F(\phi)\lambda=0.
  \label{ElkoMotionEq2Yukawa}
\end{eqnarray}
Here $\hat{g}_{\mu\nu}$ is the induced metric on the brane, and $\hat{\mathfrak{D}}_{\mu}\lambda=(\partial_{\mu}+\hat{\Omega}_{\mu})\lambda$ with $\hat{\Omega}_{\mu}$ the spin connection constructed by the induced metric $\hat{g}_{\mu\nu}$. {From $\hat{\mathfrak{D}}_{\mu}\hat{e}^{a}_{~\nu}=0$,} we can get $\hat{\mathfrak{D}}_{\mu}\hat{g}^{\lambda\rho}=\hat{\mathfrak{D}}_{\mu}(\hat{e}_{a}^{~\lambda}\hat{e}^{a\rho})=0$. Thus, the above equation can be simplified as
\begin{eqnarray}
 \frac{1}{\sqrt{-\hat{g}}}\hat{\mathfrak{D}}_{\mu}(\sqrt{-\hat{g}}\hat{g}^{\mu\nu}\hat{\mathfrak{D}}_{\nu}\lambda)
  -\! A'\gamma^{5}\gamma^{\mu} \hat{\mathfrak{D}}_{\mu}\lambda
   \!-\! {A'}^{2}\lambda
  \!+\!\text{e}^{-3A}\partial_{z}(\text{e}^{3A}\partial_{z}\lambda)-2\eta\text{e}^{2A} F(\phi)\lambda=0.~~
  \label{Elko's motion equation for brane Yukawa}
\end{eqnarray}
{Next, we introduce the following KK decomposition}
\begin{eqnarray}
\lambda_{\pm}&=&\text{e}^{-3A/2}\sum_{n}\left(\alpha_{n}(z)\varsigma^{(n)}_{\pm}(x)+\alpha_{n}(z)\tau^{(n)}_{\pm}(x)\right)\nonumber\\
             &=&\text{e}^{-3A/2}\sum_{n}\alpha_{n}(z)\hat{\lambda}_{\pm}^{n}(x).\label{KK decomposition Yukawa}
\end{eqnarray}
{For simplicity, we omit the $\pm$ subscript for the $\alpha_n$ functions in the following.} $\varsigma^{n}_{\pm}(x)$ and $\tau^{(n)}_{\pm}(x)$ are linear independant {four-dimensional} Elko spinors{, and they} satisfy
\begin{eqnarray}
\gamma^{\mu}\hat{\mathfrak{D}}_{\mu}\varsigma_{\pm}(x)=\mp\text{i}\varsigma_{\mp}(x),~\gamma^{\mu}\hat{\mathfrak{D}}_{\mu}\tau_{\pm}(x)&=&\pm\text{i}\tau_{\mp}(x),\\
\gamma^5\varsigma_{\pm}(x)=\pm\tau_{\mp}(x),~~~~~~\gamma^5\tau_{\pm}(x)&=&\mp\varsigma_{\mp}(x).
\end{eqnarray}
And the {four-dimensional} Elko spinor {$\hat{\lambda}^{n}$} should satisfy the K-G equation:
\begin{eqnarray}
\frac{1}{\sqrt{-\hat{g}}}\hat{\mathfrak{D}}_{\mu}(\sqrt{-\hat{g}}\hat{g}^{\mu\nu}\hat{\mathfrak{D}}_{\nu}\hat\lambda^n)=m^2_n\hat\lambda^n
\end{eqnarray}
with $m_n$ the mass of the Elko {spinor} on the brane. Thus, we can get the following equations of motion for the Elko KK modes $\alpha_{n}$:
\begin{eqnarray}
\alpha_{n}''-\big(\frac{3}{2}A''+\frac{13}{4}(A')^{2}+2\eta\text{e}^{2A}F(\phi)-m_{n}^2+\text{i}m_{n}A'\big)\alpha_{n}=0.\label{KKequationforElkoYukawa}
\end{eqnarray}
For the purpose of getting the action of the {four-dimensional} massless and massive Elko {spinors} from the action of a five-dimensional {massless} Elko {spinor with Yukawa-like coupling}:
\begin{eqnarray}
S_{\text{Elko}}&=&\int d^5x\sqrt{-g}\bigg[-\frac{1}{4}g^{MN}(\mathfrak{D}_M\mathop \lambda\limits^\neg\mathfrak{D}_{N}\lambda+\mathfrak{D}_N\mathop \lambda\limits^\neg\mathfrak{D}_{M}\lambda)-\eta F(\phi)\mathop \lambda\limits^\neg\lambda\bigg]\nonumber\\
               &=&-\frac{1}{2}\sum_{n}\int d^4x\bigg[\frac{1}{2}\hat{g}^{\mu\nu}(\hat{\mathfrak{D}_\mu}\hat{\mathop \lambda\limits^\neg}^{n}\hat{\mathfrak{D}_{\nu}}\hat{\lambda}^{n}+\hat{\mathfrak{D}_\nu}\hat{\mathop \lambda\limits^\neg}^{n}\hat{\mathfrak{D}_{\mu}}\hat{\lambda}^{n})+m_{n}^2\hat{{\mathop \lambda\limits^\neg}}^{n}\hat{\lambda}^{n}\bigg],
\end{eqnarray}
we should introduce the following orthonormality condition for $\alpha_{n}$:
{
\begin{eqnarray}
\int \alpha^{*}_{n}\alpha_{m}dz&=&\delta_{nm}.\label{orthonormality relation 1}
\end{eqnarray}}

{For the {Elko zero mode} ($m_0=0$),} Eq. (\ref{KKequationforElkoYukawa}) reads
\begin{eqnarray}
[-\partial_{z}^{2}+V^{Y}_{0}(z)]\alpha_{0}(z)=0, \label{Schrodinger equation 2}
\end{eqnarray}
where
\begin{eqnarray}
V^{Y}_{0}(z)=\frac{3}{2}A''+\frac{13}{4}{A'}^{2}+2\eta \text{e}^{2A}F(\phi). \label{effective potential Vz 2}
\end{eqnarray}
{For this case}, the orthonormality condition is given by
\begin{eqnarray}
\int \alpha^*_0\alpha_0dz=1.\label{orthonormality relation for zero mode}
\end{eqnarray}
As we show in {our previous work} \cite{Liu:2011nb}, there exist many similarities between the Elko field and the scalar field. {For a five-dimensional free massless scalar field,} the Schr\"{o}dinger-like equation for the {scalar zero mode} $h_0$ \cite{Liu:2008wd,Liu:2011nb} can {be read as}
\begin{eqnarray}
[-\partial_{z}^{2}+V_{\Phi}]h_0&=&[-\partial_{z}^{2}+\frac{3}{2}A''+\frac{9}{4}{A'}^{2}]h_0\nonumber\\
&=&\left[\partial_{z}+\frac{3}{2}A'\right]\left[-\partial_{z}+\frac{3}{2}A'\right]h_{0}\nonumber\\
&=&0.
\end{eqnarray}
{The solution is given by} $h_{0}(z)\varpropto\text{e}^{\frac{3}{2}A(z)}$ and it {satisfies} the orthonormality relation for any brane embedded in a five-dimensional Anti-de Sitter (AdS) {spacetime}. However, the effective potential $V_0$ for the five-dimensional {free} massless Elko {spinor} is \cite{Liu:2011nb}
\begin{eqnarray}
V_0(z)=\frac{3}{2}A''+\frac{13}{4}A'^2=\frac{3}{2}A''+\frac{9}{4}A'^2+A'^2.
\end{eqnarray}
The {additional} term $A'^2$ prevents {the localization of the zero mode}.

When the Yukawa-like coupling is introduced, {the coefficient numbers of $A''$ and $A'^2$ can be regulated}:
\begin{eqnarray}
\frac{3}{2}A''+\frac{13}{4}{A'}^{2}+2\eta \text{e}^{2A}F(\phi)=(p A')'+(p A')^2, \label{Fphiequation}
\end{eqnarray}
where $p$ is a real constant. From Eq. (\ref{Fphiequation}), the form of $F(\phi)$ can be got
\begin{eqnarray}
F(\phi)=-\frac{1}{2\eta}\text{e}^{-2A}\left[\left(p-\frac{3}{2}\right)A''+\left(p^2-\frac{13}{4}\right)A'^2\right].~~~~\label{Yukawa}
\end{eqnarray}
{Then,} Eq.~(\ref{Schrodinger equation 2}) can be rewritten as
\begin{eqnarray}
[-\partial_{z}^{2}+V^{Y}_{0}]\alpha_{0}
&=&[-\partial_{z}^{2}+p A''+p^2A'^2]\alpha_{0}\nonumber\\
&=&\left[\partial_{z}+p A'\right]\left[-\partial_{z}+p A'\right]\alpha_{0}\nonumber\\
&=&0,
\end{eqnarray}
and the Elko zero mode reads
\begin{eqnarray}
\alpha_0(z)\varpropto\text{e}^{p A(z)}. \label{Yukawazeromode}
\end{eqnarray}

{\subsection{Non-minimal coupling}}
On the other hand, for the {non-minimal} coupling, the action could be written as
\begin{eqnarray}
S&=&\int d^5x\sqrt{-g}f(\phi)\mathfrak{L}_{\text{Elko}},\nonumber \\
\mathfrak{L}_{\text{Elko}}&=&-\frac{1}{4}g^{MN}\left(\mathfrak{D}_{M}\mathop \lambda\limits^\neg\mathfrak{D}_{N}\lambda+\mathfrak{D}_{N}\mathop \lambda\limits^\neg\mathfrak{D}_{M}\lambda\right).~~~~\label{Lagrangian for Elko in 5D f phi}
\end{eqnarray}
Here $f(\phi)$ is {a function} of the background scalar field $\phi$, {which is only a} function of the extra dimension $z$ (or $y$).
From the action (\ref{Lagrangian for Elko in 5D f phi}) the following {equation of motion} can be got
\begin{eqnarray}
\frac{1}{\sqrt{-g}f(\phi)}\mathfrak{D}_{M}(\sqrt{-g}f(\phi)g^{MN}\mathfrak{D}_{N}\lambda)=0. \label{Elko's motion equation}
\end{eqnarray}
By considering {the metric} (\ref{conformally flat line-element}) and using the non-vanishing components of the spin connection (\ref{spin connection for brane}), we can rewrite Eq. (\ref{Elko's motion equation}) as:
{
\begin{eqnarray}
  &&\frac{1}{\sqrt{-\hat{g}}}\hat{\mathfrak{D}}_{\mu}(\sqrt{-\hat{g}}\hat{g}^{\mu\nu}\hat{\mathfrak{D}}_{\nu}\lambda)
 +\bigg[-\frac{1}{4}{A'}^{2}\hat{g}^{\mu\nu}\gamma_{\mu}\gamma_{\nu}\lambda\nonumber\\
 &+&\frac{1}{2}A'
 \Big( \frac{1}{\sqrt{-\hat{g}}} \hat{\mathfrak{D}}_{\mu}
       ( {\sqrt{-\hat{g}}} \hat{g}^{\mu\nu}\gamma_{\nu}\gamma_{5}\lambda)
 +\hat{g}^{\mu\nu}\gamma_{\mu}\gamma_{5}\hat{\mathfrak{D}}_{\nu}\lambda\Big)
 \nonumber\\
 &+&\text{e}^{-3A}f^{-1}(\phi)\partial_{z}(\text{e}^{3A}f(\phi)\partial_{z}\lambda)\bigg]\nonumber\\
 &=&\frac{1}{\sqrt{-\hat{g}}}\hat{\mathfrak{D}}_{\mu}(\sqrt{-\hat{g}}\hat{g}^{\mu\nu}\hat{\mathfrak{D}}_{\nu}\lambda)
  -\! A'\gamma^{5}\gamma^{\mu} \hat{\mathfrak{D}}_{\mu}\lambda
   \!-\! {A'}^{2}\lambda\nonumber\\
   &\!+\!&\text{e}^{-3A}f^{-1}(\phi)\partial_{z}(\text{e}^{3A}f(\phi)\partial_{z}\lambda)=0.
  \label{ElkoMotionEq2}
\end{eqnarray}}
In this case, we introduce the following KK decomposition:
{
\begin{eqnarray}
\lambda_{\pm}&=&\text{e}^{-3A/2}f(\phi)^{-\frac{1}{2}}\sum_{n}\left(\alpha_{n}(z)\varsigma^{(n)}_{\pm}(x)+\alpha_{n}(z)\tau^{(n)}_{\pm}(x)\right)\nonumber\\
             &=&\text{e}^{-3A/2}f(\phi)^{-\frac{1}{2}}
             \sum_{n}\alpha_{n}(z)\hat{\lambda}_{\pm}^{n}(x).\label{Elkodecomposition}
\end{eqnarray}}
By noticing {the} linear independance of {the}  $\varsigma^{(n)}_{+}$ and $\tau^{(n)}_{+}$ ($\varsigma^{(n)}_{-}$ and $\tau^{(n)}_{-}$) and the K-G equation {of the four-dimenional Elko spinor}, the {equation of motion of} the KK mode $\alpha_{n}$ can be {got}
\begin{eqnarray}
\alpha_{n}''-\big(&-&\frac{1}{4}f^{-2}(\phi)f'^{2}(\phi)+\frac{3}{2}A'f^{-1}(\phi)f'(\phi)+\frac{1}{2}f^{-1}(\phi)f''(\phi)\nonumber\\
&+&\frac{3}{2}A''+\frac{13}{4}(A')^{2}-m_{n}^2+\text{i}m_{n}A'\big)\alpha_{n}=0.\label{KKequationforElko}
\end{eqnarray}
{For} the Yukawa-like coupling case, by introducing the {orthonormality conditions (\ref{orthonormality relation 1})}, we can get the action of the four-dimenional massless and massive Elko {spinors} from the action (\ref{Lagrangian for Elko in 5D f phi}):
\begin{eqnarray}
S_{\text{Elko}}&=&-\frac{1}{4}\int d^5x\sqrt{-g}f(\phi)g^{MN}(\mathfrak{D}_M\mathop \lambda\limits^\neg\mathfrak{D}_{N}\lambda+\mathfrak{D}_N\mathop \lambda\limits^\neg\mathfrak{D}_{M}\lambda)\nonumber\\
               &=&-\frac{1}{2}\sum_{n}\int d^4x\bigg[\frac{1}{2}\hat{g}^{\mu\nu}(\hat{\mathfrak{D}_\mu}\hat{\mathop \lambda\limits^\neg}^{n}\hat{\mathfrak{D}_{\nu}}\hat{\lambda}^{n}+\hat{\mathfrak{D}_\nu}\hat{\mathop \lambda\limits^\neg}^{n}\hat{\mathfrak{D}_{\mu}}\hat{\lambda}^{n})+m_{n}^2\hat{{\mathop \lambda\limits^\neg}}^{n}\hat{\lambda}^{n}\bigg].
\end{eqnarray}

{For the Elko zero mode with $m_n=0$, Eq. (\ref{KKequationforElko}) is} simplified as
\begin{eqnarray}
[-\partial_{z}^{2}+V^{N}_{0}(z)]\alpha_{0}(z)=0, \label{KKequationforzeromode}
\end{eqnarray}
where the effective potential $V^{N}_{0}$  is given by
\begin{eqnarray}
V^{N}_{0}(z)=-\frac{1}{4}f^{-2}(\phi)f'^{2}(\phi)+\frac{3}{2}A'f^{-1}(\phi)f'(\phi)+\frac{1}{2}f^{-1}(\phi)f''(\phi)+\frac{3}{2}A''+\frac{13}{4}A'^2, \label{effective potential Vz}
\end{eqnarray}
and {the Elko zero mode $\alpha_{0}(z)$ satisfies the orthonormality condition}  (\ref{orthonormality relation for zero mode}).
By introducing three new {functions} $B(z)$, $C(z)$ and $D(z)$ satisfying
\begin{eqnarray}
B(z)&=&\frac{f'(\phi)}{f(\phi)} 
    =-3A'+\frac{A'^2}{C}-C-\frac{C'}{C},\\
\partial_zD(z)&=&\frac{3}{2}A'+\frac{1}{2}B+C,
\end{eqnarray}
the effective potential (\ref{effective potential Vz}) reads
\begin{eqnarray}
V^{N}_0(z)&=&\frac{1}{4}B^2+\frac{3}{2}A'B+\frac{1}{2}B'+\frac{3}{2}A''+\frac{13}{4}A'^2\nonumber\\
          &=&D''+D'^2,\label{effective potential VzB}
\end{eqnarray}
{and Eq.} (\ref{KKequationforzeromode}) can be reduced as
\begin{eqnarray}
[-\partial_{z}^{2}+V^N_{0}]\alpha_{0}
&=&[-\partial_{z}^{2}+D''+D'^2]\alpha_{0}\nonumber\\
&=&\left[\partial_{z}+D'\right]\left[-\partial_{z}+D'\right]\alpha_{0}\nonumber\\
&=&0.
\end{eqnarray}
In addition, it will be convenient to define a new function $K(z)$:
\begin{eqnarray}
K(z)\equiv\frac{C'}{C}-C \label{K}.
\end{eqnarray}
It should be noticed that the form of $K(z)$ is arbitrary, and the forms of $C(z)$ and $B(z)$ are determined by the warp factor and any given $K(z)$. {Now} it is easy to get {the zero mode} $\alpha_0$:
\begin{eqnarray}
 \alpha_{0}(z)&\varpropto& \text{e}^{D(z)}\nonumber\\
   &=&{\exp} \left[\frac{1}{2}\int^{z}_{0}\left(\frac{A'^2}{C}-\frac{C'}{C}+ C \right)d\bar{z}\right]\nonumber\\
    &=&{\exp}\left[\frac{1}{2}\int^{z}_{0}\left(\frac{A'^2}{C}-K\right)d\bar{z}\right]\nonumber\\
   &=&{\exp}\left[\frac{1}{2}\int^{z}_{0}\frac{A'^2}{C}d\bar{z}\right]{\exp}\left[-\frac{1}{2}\int^{z}_{0}Kd\bar{z}\right],\label{zeromodeforz2}
\end{eqnarray}
and the form of $f(\phi)$:
\begin{eqnarray}
f(\phi(z))&=&\text{e}^{\int^{z}_{0} B(\bar{z}) d\bar{z}}\nonumber\\
          &=&{\exp}\left[\int^{z}_{0}\left( -3A'+\frac{A'^2}{C}- C-\frac{C'}{C} \right) d\bar{z}\right]\nonumber\\
           &=&{\exp}\left[-3A-2\ln |C|+\int^{z}_{0}\left(\frac{A'^2}{C}+K \right)d\bar{z}\right]\nonumber\\
          &=&\text{e}^{-3A}C^{-2}{\exp}\left[\int^{z}_{0}\frac{A'^2}{C}d\bar{z}\right]{\exp}\left[\int^{z}_{0}Kd\bar{z}\right].~~~~~\label{fphi}
\end{eqnarray}
It is not difficult to check that the orthonormality condition (\ref{orthonormality relation for zero mode}) always requires that $K(z)$ is an odd function and positive as $z>0$.
Here Eqs.~(\ref{zeromodeforz2}) and (\ref{fphi}) give the general expressions of the zero mode $\alpha_0$ and the function $f(\phi)$ because the function $C(z)$ can be expressed by the function $K(z)$ according to Eq.~(\ref{K}):
\begin{eqnarray}
C(z)=\frac{\text{e}^{\int^z_1 K(\bar{z})d\bar{z}}}{\mathcal{C}_1-\int_1^z\text{e}^{\int^{\hat{z}}_1 K(\bar{z})d\bar{z}}d\hat{z}}, \label{Cz}
\end{eqnarray}
{where $\mathcal{C}_1$ is an arbitrary parameter.}
As we {showed} in our previous work, the role of $K(z)$ is similar to the auxiliary superpotential $W(\phi)$, which is introduced in order to solve the Einstein equations in thick brane models. For a given $K(z)$ the zero mode $\alpha_0$ is obtained by integrating Eq. (\ref{zeromodeforz2}). Then, the scalar field function $f(\phi(z))$ is determined by integrating Eq. (\ref{fphi}). For different forms of $K(z)$, there exist different configurations of {the zero mode} $\alpha_0$ and function $f(\phi)$. It {gives} us more choices and possibilities to {study} the localization of {the Elko zero mode} on the branes. Next we will consider the localization of {the Elko zero mode with} this two kinds of couplings on dS/AdS thick branes.

\section{Localization of Elko zero mode on dS/AdS thick branes }
\label{section3}
In this section, we investigate the localization of {the Elko zero mode with} two kinds of couplings on single-scalar-field generated dS/AdS thick {branes}~\cite{Afonso:2006gi,Zhang:2016ksq}. The system is described by the action
\begin{eqnarray}
S=\int{d^5x\sqrt{-g}~\left[\frac{M_5}{4}R-\frac{1}{2}\partial_M\phi\partial^M\phi-V(\phi)\right]},
\end{eqnarray}
where $R$ is the five-dimensional scalar curvature and $V(\phi)$ is the potential {of the} scalar field. For convenience, {the} fundamental mass {scale $M_5$ is set to} 1. The line element is described by Eq. (\ref{the line-element for 5D space time}) and the induced metrics $\hat{g}_{\mu\nu}$ on the branes read
\begin{eqnarray}
    \hat{g}_{\mu\nu}=\left\{
    \begin{array}{cc}
    -dt^2+e^{2\beta t}(dx_1^2+dx_2^2+dx_3^2)~~~\textrm{$\text{dS}_4$ brane},\label{ds}\\
    e^{-2\beta x_3}(-dt^2+dx_1^2+dx_2^2)+dx_3^2~~~\textrm{$\text{AdS}_4$ brane}
    \label{Ads}.
    \end{array}\right.~~
\end{eqnarray}
Here the parameter $\beta$ is related to the the four-dimensional cosmological constant of the dS$_4$ or AdS$_4$ brane by $\Lambda_4=3\beta^2$ or $\Lambda_4=-3\beta^2$ \cite{Liu:2009dt,Liu:2011zy,Zhang:2016ksq}. By introducing the following {scalar potential}
\begin{eqnarray}
 V(\phi)&=&\frac{3}{4}a^2(1+\Lambda_{4}) \big[1+(1+3s)\Lambda_{4} \big]\cosh^2(b\phi)\nonumber\\
             &&-3a^2(1+\Lambda_{4})^2\sinh^2(b\phi),\label{phi}
\end{eqnarray}
a brane solution can be obtained~\cite{Afonso:2006gi,Zhang:2016ksq}:
\begin{eqnarray}
    A(y)&=&-\frac{1}{2}\ln[sa^2(1+\Lambda_{4})\sec^2 \bar{y} ],\label{warpfactory}\\
    \phi(y)&=&\frac{1}{b}\text{arcsinh}(\tan \bar{y} ),\label{scalar}
\end{eqnarray}
where $\bar{y}\equiv a(1+\Lambda_{4})y$. {The parameters $a$, $s$ and $b$ are real} with $s\in(0,1]$ and $b=\sqrt{\frac{2(1+\Lambda_{4})}{3(1+(1+s)\Lambda_{4})}}$. Note that the thick brane is extended in the range  $y\in\left(-|\frac{\pi}{2a(1+\Lambda_{4})}|,~|\frac{\pi}{2a(1+\Lambda_{4})}|\right)$. By performing the coordinate transformation (\ref{coordinate transformation}), we can get
    \begin{eqnarray}
    y=\frac{1}{a(1+\Lambda_{4} )}
       \left[2\arctan\left(e^{hz}\right)
             -\frac{\pi }{2}
       \right]              \label{yz}
    \end{eqnarray}
{with} $h\equiv\sqrt{\frac{1+\Lambda_{4} }{s}}$.
It should be noticed that the range of {the coordinate} $z$ will trend to {infinite}.
By substituting the relation (\ref{yz}) into {the solution (\ref{warpfactory}) and (\ref{scalar})}, we can obtain the warp factor and scalar field in the coordinate $z$ \cite{Zhang:2016ksq}:
    \begin{eqnarray}
    A(z)&=&-\frac{1}{2} \ln\big[a^2 s (1+\Lambda_{4} ) \cosh^2(hz)\big], \label{warpfactorscalarz} \\
    \phi(z)&=&\frac{1}{b}\text{arcsinh} \left[\sinh(hz)\right]
        =\frac{h}{b}z.\label{phi}
    \end{eqnarray}
The warp factor $\text{e}^{2A(z)}$ {is convergent} at boundary. When $\Lambda_{4}=0$, the above solution {reduces to the flat brane one.}

\subsection{Yukawa-like coupling}

\begin{figure*}[!htb]
    \includegraphics[width=0.45\textwidth]{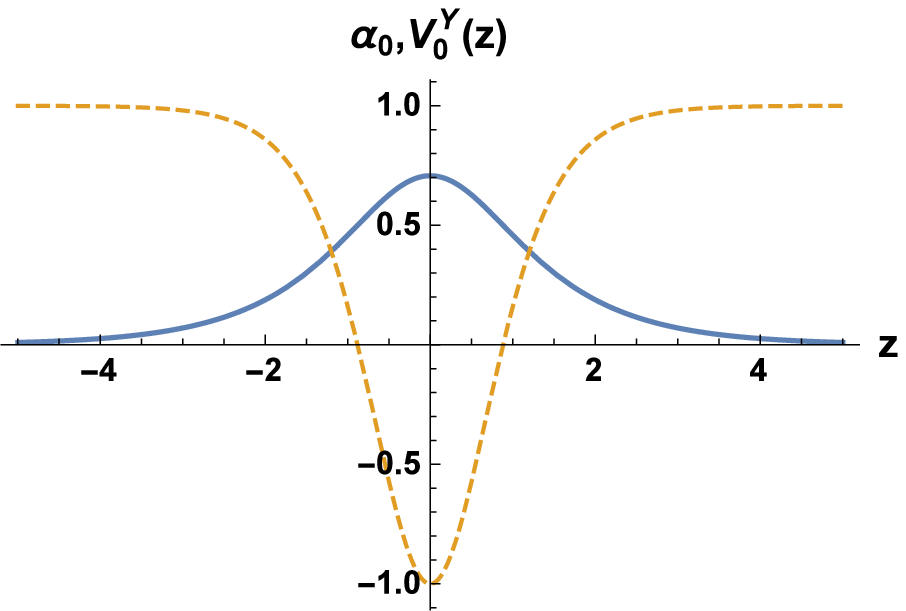}
    \includegraphics[width=0.45\textwidth]{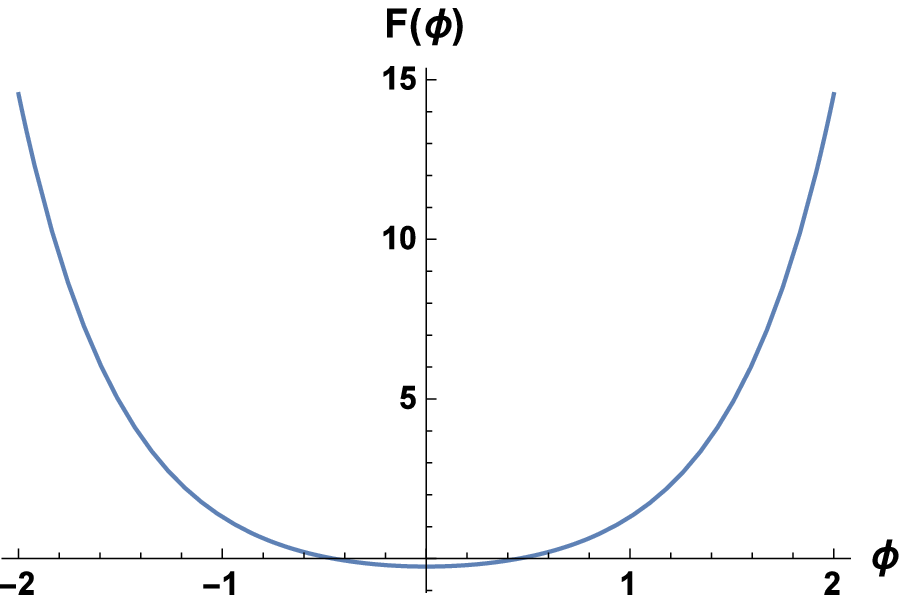}
    \vskip 1mm \caption{The shapes of
the Elko zero mode $\alpha_{0}(z)$ (\ref{zeromode1z}) (thick line), the effective potential $V^Y_0(z)$ (dashed line) are drawn on the left and the shape of function $F(\phi)$ (\ref{Fphi1z}) (right) is drawn on the right. The parameters are set to $h=b=p=\eta=a^2s(1 + \Lambda_4)=1$. }
    \label{z1}
\end{figure*}

Firstly, we consider the Yukawa-like coupling mechanism for the localization of the Elko zero mode.
According to Eqs. (\ref{Fphiequation})-(\ref{Yukawazeromode}), (\ref{warpfactorscalarz}) and (\ref{phi}), the Elko zero mode,  the function $F(\phi)$ and the effective potential $V_0^{Y}$ are given by
\begin{eqnarray}
\alpha_0&\propto&\text{e}^{pA(z)}
       =(a^2s(1+\Lambda_4))^{-\frac{p}{2}}\text{sech}^p(hz),\label{zeromode1z}\\
F(\phi)
       &=&-\frac{h^2a^2s}{16\eta}(1+\Lambda_4)\left[25-4p(2+p)+(4p^2-13)\cosh(2b\phi)\right],~~\label{Fphi1z} \\
V_0^Y &=&  p A'' + p^2A'^2 =-p h^2\text{sech}^2(hz)+p^2h^2\tanh^2(hz).      \label{V0Yz}
\end{eqnarray}
Now the orthonormality condition reads
\begin{eqnarray}
&&\int \alpha^*_0\alpha_0dz=\int \alpha_0^2dz\nonumber\\
&\propto&\int (a^2s(1+\Lambda_4))^{-p}\text{sech}^{2p}(hz)dz\nonumber\\
&=&\frac{4^p}{hp}(a^2s(1+\Lambda_4))^{-p}{}_{2}\text{F}_{1}\left(p,2p;1+p;-1\right)<\infty. \label{orthonormality relation for z1}
\end{eqnarray}
It requires $p>0$. In Fig. \ref{z1} we plot the {shapes} of the zero mode, the effective potential $V_0^Y(z)$ and the function $F(\phi)$, which {show} that the Elko zero mode can be localized on the brane.
Therefore, the Yukawa-like coupling mechanism can be successfully used to localize the zero mode of the Elko spinor on the dS/AdS thick brane.{ We can find that the effective potential $V_0^Y(z)$ is a PT potential. The shape of $F(\phi)$} has a {minimum} around $\phi=0$ and diverges when $\phi\rightarrow\infty$. {As $\phi\rightarrow\infty$} the boundary values of {the} warp factor $\text{e}^{2A}$ and $F(\phi)$ are just opposite because there exists a factor $\text{e}^{-2A}$ in the expression of $F(\phi)$ (\ref{Yukawa}).

\subsection{Non-minimal coupling}

Next, we focus on the {non-minimal} coupling {mechanism}. {As shown} in our previous work \cite{Zhou:2017bbj}, {there exist different configurations of the Elko zero mode and $f(\phi)$ for different choices of $K(z)$.} In this paper we will consider two kinds of $K(z)$ and investigate the localization of the Elko zero mode (\ref{zeromodeforz2}).

\subsubsection{$K(z)=-kA'$}
\begin{figure*}[!htb]
    \includegraphics[width=0.45\textwidth]{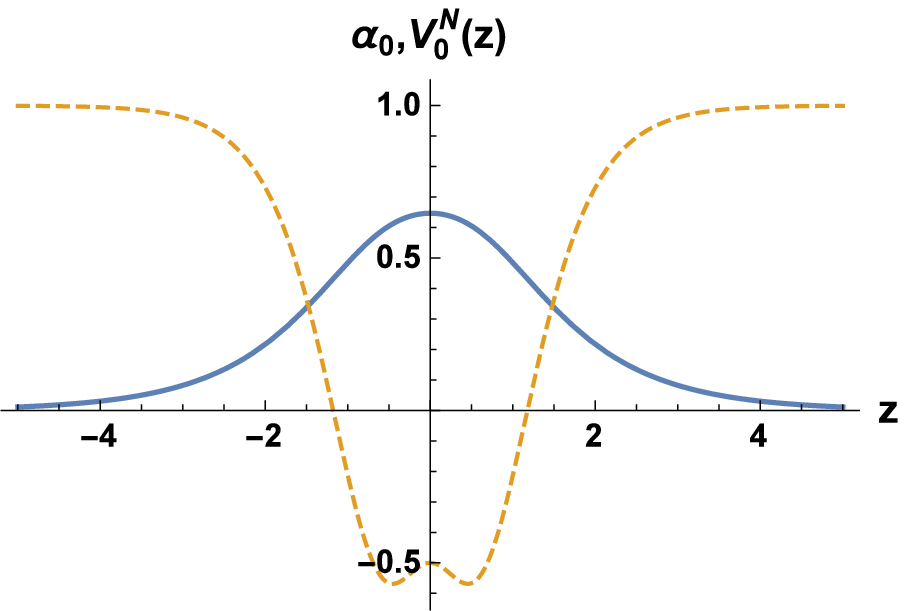}
    \includegraphics[width=0.45\textwidth]{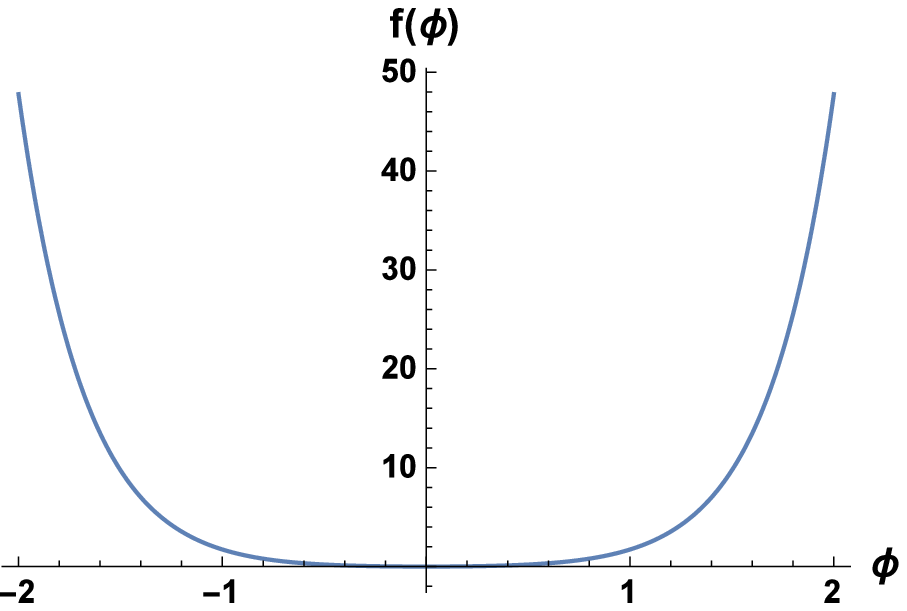}
    \vskip 1mm \caption{{The shapes of the Elko zero mode $\alpha_{0}(z)$ (\ref{zeromode2}) (thick line), the effective potential $V_0^N(z)$ (\ref{V0N1}) (dashed line) are drawn on the left and the shape of function $f(\phi)$ (\ref{fphi1}) (right) is drawn on the right.} The parameters are set to $h=b=a^2s(1 + \Lambda_4)=1$. }
    \label{z2}
\end{figure*}

Firstly, {it is} a natural choice to consider $K(z)=-kA'$ with $k$ a positive constant. It is easy to get
\begin{eqnarray}
C(z)=\frac{1}{\mathcal{C}_1\cosh(hz)^{-k}-\frac{\coth(hz)\mathcal{F}(z)\sqrt{-\sinh^2(hz)}}{h+hk}}.
\end{eqnarray}
Here $\mathcal{F}(z)={}_{2}\text{F}_{1}\left(\frac{1}{2},\frac{1+k}{2};\frac{3+k}{2};\cosh^2(hz)\right)$. Especially, when $k=1$, the form of $C(z)$ will {be reduced} to
\begin{eqnarray}
C(z)=\frac{h\cosh(h z)}{h\mathcal{C}_1-\sinh(hz)}.
\end{eqnarray}
Here $\mathcal{C}_1$ is an arbitrary constant and we always let $\mathcal{C}_1=0$. Thus, the zero mode is rewritten as
\begin{eqnarray}
 \alpha_{0}(z)&\varpropto& \text{e}^{D(z)}\nonumber\\
 &=&{\exp}\left[\frac{1}{2}\int^{z}_{0}\frac{A'^2}{C}d\bar{z}\right]{\exp}\left[-\frac{1}{2}\int^{z}_{0}Kd\bar{z}\right]\nonumber\\
 &\varpropto&\text{sech}(hz){\exp}\left[-\frac{1}{4}\text{sech}^2(hz)\right].  \label{zeromode2}
\end{eqnarray}
It is easy to check that the orthonormality condition
\begin{eqnarray}
&&\int \alpha^*_0\alpha_0dz=\int \alpha_0^2dz\nonumber\\
&\varpropto&\int \text{sech}^2(h\bar{z}){\exp}[-\frac{1}{2}\text{sech}^2(h\bar{z})] dz \nonumber\\
&=&2\sqrt{2}F({\sqrt{2}}/{2})<\infty\label{orthonormality relation for z2}
\end{eqnarray}
can be satisfied. Here, $F(z)$ gives the Dawson integral and $F({\sqrt{2}}/{2})=0.512496$. {For this case,} the effective potential $V_0^N(z)$ is given by
\begin{eqnarray}
V_0^N(z)=\frac{h^2}{32}\left(-2-5\cosh(2hz)-10\cosh(4hz)+\cosh(6hz)\right)\text{sech}^6(hz).\label{V0N1}
\end{eqnarray}
We plot the shapes of the zero mode and the effective potential in Fig. \ref{z2}, from which we can see that the zero mode (\ref{zeromode2}) is localized on the brane {and the effective potential is a PT-like potential}. The function $f(\phi)$ reads
\begin{eqnarray}
f(\phi(z)) 
          =\frac{a^4s^2}{h^2}(1 + \Lambda_4)^2\cosh^3(b\phi)\tanh^2(b\phi)~{\exp}\left[-\frac{1}{2}\text{sech}^2(b\phi)\right], \label{fphi1}
\end{eqnarray}
{and it is plotted} in Fig.~(\ref{z2}) with $h=b=a^2s(1 + \Lambda_4)=1$. It is obvious that the shape of $f(\phi)$ is similar to $F(\phi)$ in the previous subsection, {and it} has a minimum at the point of $\phi=0$ and diverges {at infinity}. The boundary {behaviour} of $f(\phi)$ is also opposite to the warp factor, because there exists {a factor }$\text{e}^{-3A}$ in the expression of $f(\phi)$ (\ref{fphi}).

\subsubsection{$K(z)=k\phi$}

\begin{figure*}[!htb]
    \includegraphics[width=0.45\textwidth]{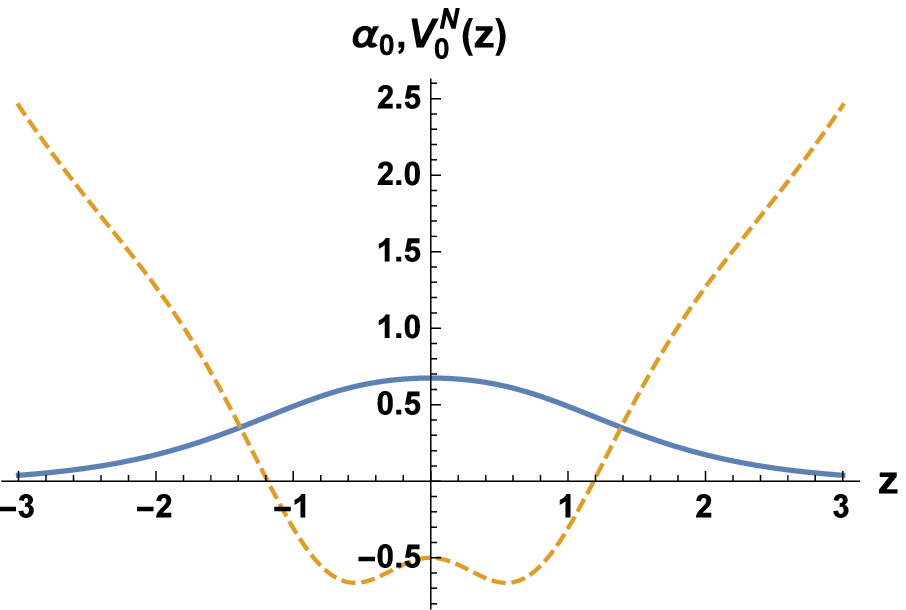}
    \includegraphics[width=0.45\textwidth]{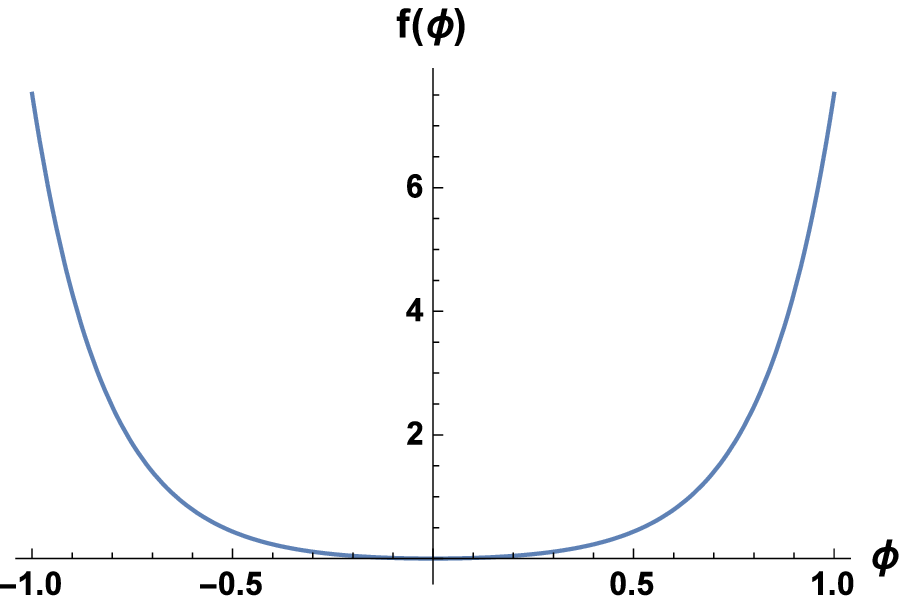}
    \vskip 1mm \caption{{The shapes of the Elko zero mode $\alpha_{0}(z)$ (\ref{zeromodez3}) (thick line), the effective potential $V_0^N(z)$ (dashed line) are drawn on the left and the shape of function $f(\phi)$ (right) is drawn on the right.} The parameters are set to $\bar{k}=h=b=a^2s(1 + \Lambda_4)=1$. }
    \label{z3}
\end{figure*}

{Another natural choice for $K(z)$} is $K(z)=k\phi=k\frac{h}{b}z$ with {positive} $k$, {for which} the form of $C(z)$ is
\begin{eqnarray}
C(z)=-\frac{\sqrt{\frac{2\bar{k}}{\pi}}}{\text{Erfi}\left(\sqrt{\frac{\bar{k}}{2}}z\right)}\text{e}^{\frac{\bar{k}}{2}z^2},
\end{eqnarray}
where $\bar{k}\equiv k\frac{h}{b}$ and $\text{Erfi}\left(\sqrt{\frac{\bar{k}}{2}}z\right)$ is the imaginary error function. Thus the zero mode reads
\begin{eqnarray}
 \alpha_{0}(z) 
 &\varpropto &{\exp}\left[-\frac{1}{2} h^2  \sqrt{\frac{\pi}{2\bar{k}}}~ \mathcal{L}(z)
                          -\frac{\bar{k}}{4}z^{2}\right]. \label{zeromodez3}
\end{eqnarray}
where
\begin{eqnarray}
 \mathcal{L}(z)\equiv  \int^{z}_{0}
           \text{Erfi}
            \left(\sqrt{\frac{\bar{k}}{2}}\bar{z}\right)
              \tanh^2(h\bar{z}){\exp}
            \left[\frac{-\bar{k} \bar{z}^{2}}{2}\right]  d\bar{z}.~~~~
\end{eqnarray}
It can be {verified} that the function $\mathcal{L}(z)$ approaches a constant {as $|z|\rightarrow\infty$}, which shows that $\alpha_0(|z|\rightarrow\infty)\varpropto{\exp}\left[-\frac{\bar{k}}{4}z^{2}\right]$ and its orthonormality condition can be satisfied. Thus, the zero mode (\ref{zeromodez3}) can be localized on the brane, see Fig.~\ref{z3}. The effective potential $V_0^N(z)$ and the corresponding function $f(\phi)$ have slightly complex forms and we only show their shapes in Fig. \ref{z3}. {Here, we can find the effective potential is an infinite deep potential instead of a PT one. However, the shape of $f(\phi)$ is still} similar to the ones in the previous two subsections. Therefore, the functions $F(\phi)$ and $f(\phi)$ have similar properties and play similar roles, although they appear in different places in the actions.

\section{Localization of Elko zero mode on AdS thick brane with divergent warp factor}
\label{section4}

In this section, we {will} consider another kind of AdS thick brane model and investigate the localization of {the Elko zero mode} with two kinds of couplings. The warp factor in previous section is convergent at the boundaries of the extra dimension, but the one in this section will be diverge, {for which} the zero mode of a five-dimensional free scalar field can not be localized on this brane \cite{Liu:2011zy}. {This difference will bring} us some different and interesting results. The action of this system reads \cite{Liu:2011zy}
\begin{equation}
 S= \int d^5 x \sqrt{-g}\left[\frac{1}{2} R-\frac{1}{2}
     g^{MN}\partial_M \phi \partial_N \phi - V(\phi) \right],
\label{action}
\end{equation}
where $R$ is the five-dimensional scalar curvature. {Note that $M_5$ is set to 2 here.} The metric is described by (\ref{conformally flat line-element}) and the induced metric is
{$\hat{g}_{\mu\nu}=e^{-2\beta x_3}(-dt^2+dx_1^2+dx_2^2)+dx_3^2$ with $\Lambda_4=-3\beta^2$}. For the {scalar} potential
\begin{eqnarray}\label{Vphi}
V(\phi)=-\frac{3(1+3\delta)\beta^{2}}{2\delta}\cosh^{2(1-\delta)}
            \left(\frac{\phi}{\phi_{0}} \right),
\end{eqnarray}
a thick AdS brane solution was given in {Refs.} \cite{Wang:2002pka,Liu:2011zy}:
\begin{eqnarray}
A(z)&=& -\delta\ln\left|\cos \left(\frac{\beta}{\delta}z\right) \right|,\label{warpfactor}
\\
\phi(z)&=&\phi_{0}\text{arcsinh}\left(\tan\left(\frac{\beta}{\delta}z\right)\right)\label{scalarsolution}
\end{eqnarray}
with
\begin{eqnarray}
\phi_{0}&\equiv&\sqrt{3\delta(\delta-1)}.
\end{eqnarray}
Here, the range of the extra dimension is $-z_{b}\leq z\leq z_{b}$ with
$z_{b}=\left|\frac{\delta\pi}{2\beta}\right|$ and the parameter $\delta$ satisfies $\delta >1$ or $\delta <0$. It {was found} that only when $\delta>1$, there {exists} a thick 3-brane which localizes at $|z|\approx0$ \cite{Wang:2002pka,Liu:2011zy}.
Thus, we only consider the case of $\delta>1$. In this case, the warp factor $\text{e}^{2A(z)}$ {is} diverge at the boundaries $z=\pm z_{b}$.

\subsection{Yukawa-like coupling}

\begin{figure*}[!htb]
    \includegraphics[width=0.45\textwidth]{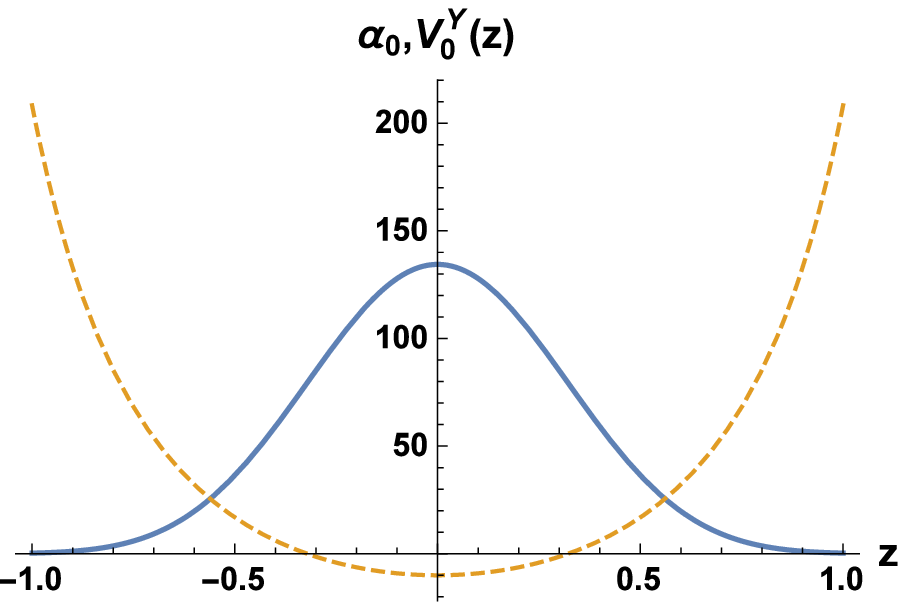}
    \includegraphics[width=0.45\textwidth]{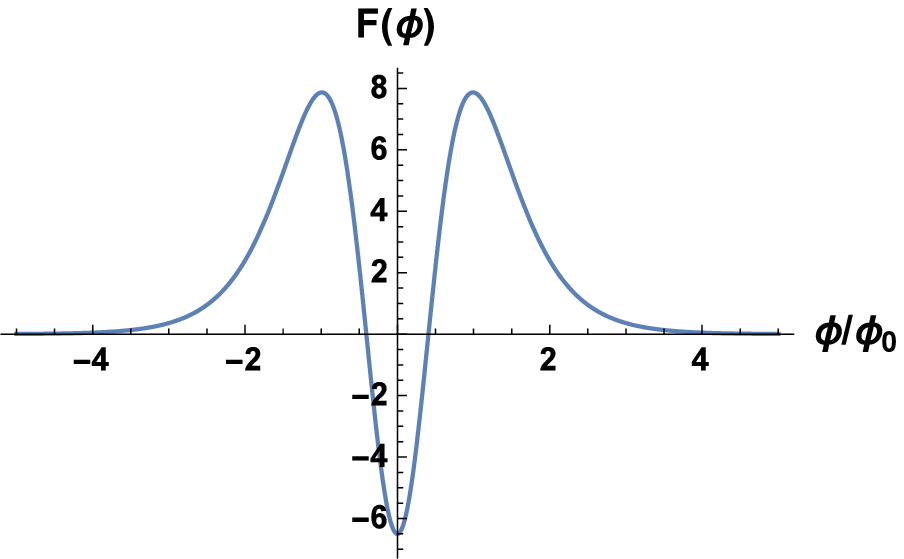}
    \vskip 1mm \caption{{The shapes of the Elko zero mode $\alpha_{0}(z)$ (\ref{zeromode4z}) (thick line), the effective potential $V^Y_0(z)$ (\ref{V0Yz2})} (dashed line) are drawn on the left and the shape of function $F(\phi)$ (\ref{Fphi2z}) (right) is drawn on the right. For visibility, here $\alpha_{0}(z)$ has been magnified 100 times. The parameters are set to $\eta=-1$, $\delta=\beta=2$, and $p=-5$. }
    \label{z4}
\end{figure*}

With the Yukawa-like coupling and by substituting Eqs. (\ref{warpfactor}) and (\ref{scalarsolution}) into (\ref{Fphiequation})-(\ref{Yukawazeromode}),the Elko zero mode $\alpha_0$, the function $F(\phi)$ and the effective potential $V_0^{Y}$ {read}
\begin{eqnarray}
\alpha_0&\propto&\text{e}^{pA(z)} = \cos^{-p\delta}\left(\frac{\beta}{\delta}z\right),\label{zeromode4z}\\
F(\phi)
            &=&\frac{\beta^2}{8\delta\beta}\left[6-4p+(6+13\delta-4p(1+p\delta))\sinh^2\left(\frac{\phi}{\phi_0}\right)\right]\cosh^{-2\delta}\left(\frac{\phi}{\phi_0}\right),~~~~~\label{Fphi2z}\\
V_0^Y &=&  p A'' + p^2A'^2=\frac{p\beta^2}{\delta}\sec^2\left(\frac{\beta}{\delta}z\right)+p^2\beta^2\tan^2\left(\frac{\beta}{\delta}z\right).  \label{V0Yz2}
\end{eqnarray}
{And} the orthonormality {condition requires $p\delta<0$:}
\begin{eqnarray}
&&\int \alpha^*_0\alpha_0dz=\int \alpha_0^2dz\nonumber\\
&\propto&\int^{\frac{\delta\pi}{2\beta}}_{-\frac{\delta\pi}{2\beta}} \cos^{-2p\delta}\left(\frac{\beta}{\delta}z\right)dz\nonumber\\
&=&\frac{\sqrt{\pi}\Gamma(\frac{1}{2}-p\delta)}{\Gamma(1-p\delta)}<\infty. \label{orthonormality relation for z4}
\end{eqnarray}
{Therefore}, the zero mode can be localized on this AdS thick brane for any negative $p$ by introducing the Yukawa-like coupling mechanism. We plot the zero mode, {the effective potential $V_0^Y(z)$ and} the function $F(\phi)$ in Fig.~\ref{z4}. { The effective potential is an infinite deep potential. And unlike previous model,} here, $F(\phi)$ has the shape of a volcano,
this is because the boundary behavior of the warp factor is changed compared to the previous section.

\subsection{{Non-minimal} coupling}

\begin{figure*}[!htb]
    \includegraphics[width=0.45\textwidth]{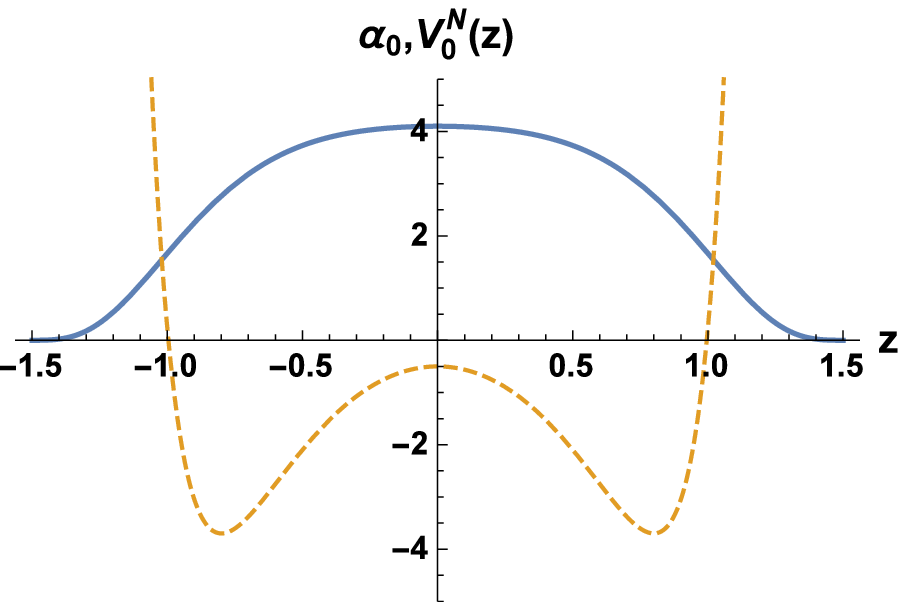}
    \includegraphics[width=0.45\textwidth]{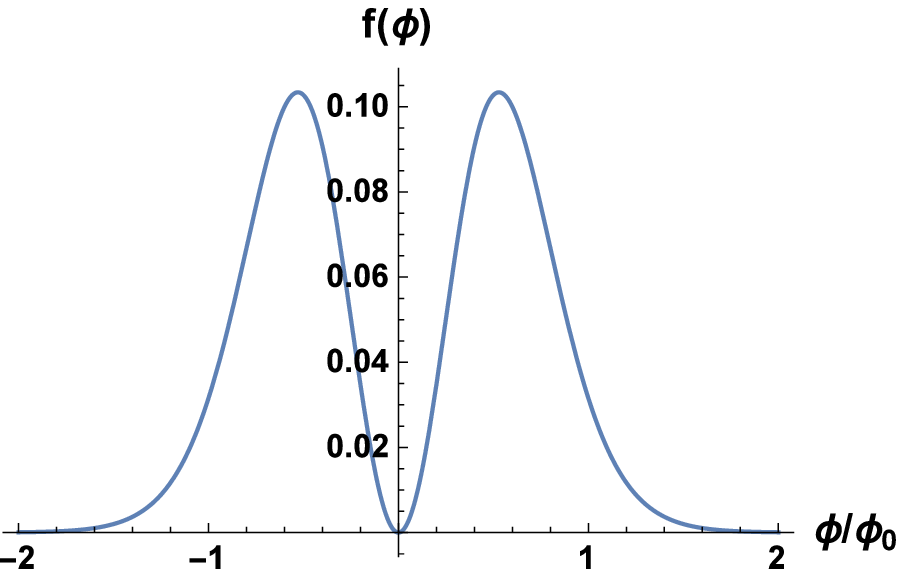}
    \vskip 1mm \caption{{The shapes of the Elko zero mode $\alpha_{0}(z)$ (\ref{zeromode5}) (thick line), the effective potential $V^N_0(z)$ (dashed line) are drawn on the left and the shape of function $f(\phi)$ (right) is drawn on the right. For visibility, here $\alpha_{0}(z)$ has been magnified 5 times.} The parameters are set to $\beta=\delta=2$. }
    \label{z5}
\end{figure*}

Finally, we turn to the {non-minimal} coupling {mechanism} {by considering} $K(z)=kA'$ with $k$ a positive constant. Note that the sign in front of $k$ is just opposite {to} the one in {the previous} section. {When} $k=\frac{1}{\delta}$, {the function} $C(z)$ in (\ref{Cz}) has the following form
\begin{eqnarray}
C(z)={ \frac{\bar{k}\sec(\bar{k}z)}
          {\ln\left(\frac{\cos\left(\frac{\bar{k}}{2}z\right)
                                   -\sin\left(\frac{\bar{k}}{2}z\right)}
                                 {\cos\left(\frac{\bar{k}}{2}z\right)
                                  +\sin\left(\frac{\bar{k}}{2}z\right)}
                      \right)} },
\end{eqnarray}
where $\bar{k}\equiv\frac{\beta}{\delta}$ {and the parameter $\mathcal{C}_1$ is set to be zero.} Then the zero mode reads
\begin{eqnarray}
\alpha_{0}(z)
 &\varpropto& \cos^{\frac{1}{2}}(\bar{k}z)~
      {\exp} \left[\frac{1}{2}\bar{k}\delta^2
             \int^{z}_{0} \sin(\bar{k}\bar{z})\tan(\bar{k}\bar{z})\ln\left(\frac{\cos\left(\frac{\bar{k}}{2}\bar{z}\right)
                                   -\sin\left(\frac{\bar{k}}{2}\bar{z}\right)}
                                 {\cos\left(\frac{\bar{k}}{2}\bar{z}\right)
                                  +\sin\left(\frac{\bar{k}}{2}\bar{z}\right)}
                      \right) d\bar{z}
          \right].  \label{zeromode5}
 \end{eqnarray}
It is easy to check that the above $\alpha_{0}(z)$ vanishes when $z\rightarrow z_{b}$ and the orthonormality condition for the Elko zero mode $\int \alpha^*_0\alpha_0dz =1$ can be satisfied. Therefore, the zero mode is localized on the brane (see Fig.~\ref{z5}). {The effective potential is an infinite deep one and the function $f(\phi)$ has a closed form but we only show their shapes in Fig.~\ref{z5}.} It {can} be found that all of the functions $F(\phi)$ and $f(\phi)$ have a {minimum} round $\phi=0$ in both thick brane models.

\section{Conclusion and discussion}
\label{section5}
In this paper, we introduced two localization mechanisms to investigate the localization of the zero mode of a five-dimensional Elko {spinor} on dS/AdS thick branes. Firstly, we reviewed the two localization mechanisms, i.e, the Yukawa-type coupling and the non-minimal coupling. It showed that in order to obtain the Elko zero mode on a brane, the form of $F(\phi)$ in the Yukawa-type coupling mechanism {is} determined by the warped factor (i.e.,  $F(\phi)=-\frac{1}{2\eta}\text{e}^{-2A}\left[\left(p-\frac{3}{2}\right)A''
+\left(p^2-\frac{13}{4}\right)A'^2\right]$), and the function $f(\phi)$ in the non-minimal coupling mechanism {is determined} by introducing {an} auxiliary function $K(z)$. Then, we considered two kinds of curved thick brane models and {investigated} the localization of the Elko zero mode with the {two kinds of} localization mechanisms.

{In the first brane model, we considered the dS/AdS thick brane generated by a single scalar field.} The results showed that for the Yukawa-type coupling, the zero mode can be localized on the brane under the condition $p>0$. It {is} interesting {to note} that the factor $\text{e}^{-2A}$ in the expression of $F(\phi)$ leads to the result that the boundary behavior of $F(\phi)$ is opposite to the warp factor $\text{e}^{2A}$. Thus the shape of $F(\phi)$ {diverges when $\phi\rightarrow\infty$} because of the convergent warp {factor}. For the non-minimal coupling mechanism, by introducing two different forms of the auxiliary function $K(z)$,
the zero mode can be confined on the brane.  And the coupling functions $f(\phi)$ with two different forms of $K(z)$ {are similar to} the $F(\phi)$ for the Yukawa-type coupling in this brane system.  Therefore, these two coupling functions have similar properties and play similar roles although they appear in different places in the {action} of five-dimensional Elko spinor. {This result may help us to explore a new localization mechanism and expand the possibility of localizing Elko {spinor}.}

Next, another AdS thick brane with divergent warp factor was considered.
For the Yukawa-type coupling case, the zero mode can be localized on this AdS thick brane for any negative $p$, which is just opposite {to} the condition in the previous thick brane model. Because of the divergent warp factor in this brane system, the shape of $F(\phi)$ seems to {be a volcano.}
For the non-minimal coupling case with the given auxiliary function $K(z)$, we obtained the localized zero mode on the brane and the coupling function $f(\phi)$. The function $f(\phi)$ still has a closed form. {An interesting result is} that both $F(\phi)$ and $f(\phi)$ always have a {minimum around $\phi=0$} no matter what the warp factor is.

In this paper, we gave the expressions of {the} coupling functions {in order to localize} the zero mode of {a} five-dimensional Elko spinor on curved thick branes. {We found} that the effective potential of the Schr\"{o}dinger-like equation satisfied by the zero mode is {a} PT potential or {an infinite deep potential}. However, it should be noticed that the equation satisfied by the massive KK modes is a complex one, which is quite different from that satisfied by the zero mode. Thus the effective potential function in this paper is not applicable for the massive KK modes. And we can not judge whether there exists a massive KK mode although the effective potential is {PT-like or infinite}. {In the future, we will investigate the localization of the massive Elko KK modes on different thick branes.}

\section*{Acknowledgments}
The authors thank Professor Yu-Xiao Liu for his kind help. This work was supported in part by the National Natural Science Foundation of China (Grant Nos. 11305095, 11647016 and 11705106), the Natural Science Foundation of Shandong Province, China (Grant No. ZR2013AQ016), and the Scientic Research Foundation of Shandong University of Science and Technology for Recruited Talents (Grant No. 2013RCJJ026).

\end{document}